\documentclass[runningheads]{llncs}

\usepackage[T1]{fontenc}

\usepackage{graphicx}
\usepackage{amsmath}
\usepackage{amssymb}
\usepackage{booktabs}
\usepackage{multirow}

\begin{document}

\title{Incomplete Multimodal Learning for Complex Brain Disorders Prediction}

\titlerunning{Incomplete Multimodal Brain Data Analysis }

\author{Reza Shirkavand\inst{1} \and
Liang Zhan\inst{1} \and
Heng Huang\inst{1} \and
Li Shen\inst{2} \and
Paul M. Thompson \inst{3}
}

\authorrunning{R. Shirkavand and et al.}

\institute{Department of Electrical and Computer Engineering, University of Pittsburgh, Pittsburgh, PA, USA \email{res182@pitt.edu} \and
Perelman School of Medicine, Philadelphia, University of Pennsylvania, PA, USA \and
Keck School of Medicine, University of Southern California, Marina del Rey, CA, USA
}

\maketitle

\begin{abstract}
Recent advancements in the acquisition of various brain data sources have created new opportunities for integrating multimodal brain data to assist in early detection of complex brain disorders.  However, current data integration approaches typically need a complete set of biomedical data modalities, which may not always be feasible, as some modalities are only available in large-scale research cohorts and are prohibitive to collect in routine clinical practice. Especially in studies of brain diseases, research cohorts may include both neuroimaging data and genetic data, but for practical clinical diagnosis, we often need to make disease predictions only based on neuroimages. As a result, it is desired to design machine learning models which can use all available data (different data could provide complementary information) during training but conduct inference using only the most common data modality. 
We propose a new incomplete multimodal data integration approach that employs transformers and generative adversarial networks to effectively exploit auxiliary modalities available during training in order to improve the performance of a unimodal model at inference. We apply our new method to predict cognitive degeneration and disease outcomes using the multimodal imaging genetic data from Alzheimer's Disease Neuroimaging Initiative (ADNI) cohort. Experimental results demonstrate that our approach outperforms the related machine learning and deep learning methods by a significant margin.


\end{abstract}

\section{Introduction}
Alzheimer’s disease (AD) is a progressive neurodegenerative disease with primary symptoms of memory loss and other cognitive deficits. Despite decades of studies on AD, effective treatments are only just emerging \cite{barthelemy2020soluble,van2022lecanemab}. Consequently, early diagnosis is extremely important for planning support strategies and improving the quality of patients' lives.

AD symptoms are usually reflected by the cognitive status. People diagnosed as having mild cognitive impairment (MCI) are at a higher risk of progressing to AD \cite{jack2018nia}. Several studies \cite{zhu2016early,schmidt2015multi,li2021predicting} have been conducted to predict the progression from MCI to AD and aim to identify predictors that can distinguish stable MCI (sMCI) from progressive or converting MCI (pMCI or cMCI). Cognitive measures that carry information about cognitive status can help in diagnosing people at risk of AD in the earlier stages of the disease.  

Many recent studies suggested that combining multiple data modalities like structure MRI, PET, genetic data, or CSF biomarkers can improve diagnostic and prognostic accuracy for AD at different stages. Multimodal learning (MML) outperforms unimodal methods \cite{huang2021multimodalproof,sun2020tcgm}, but the acquisition cost of some modalities, particularly genetic data, limits the application of general MML models in AD studies and other medical problems. Exploiting multiple modalities in training, but only using one principal modality, \emph{e.g.} MRI - which is the most common neuroimaging modality for studying AD - at inference would be a practical workaround. Consequently, there is a strong motivation to develop incomplete multimodal learning methods that can deal with missing or partial data modalities to boost prediction performance.

Over the past decade, deep learning has been successfully used to tackle various problems in image, audio, and text domains \cite{goodfellow2016deep}. This tremendous success has been due to the inductive biases of deep neural networks taking advantage of spatial, temporal, or semantic structures in the data.  However, in the absence of such structure in tabular data (biomedical data are often processed with domain knowledge to extract features, such as the summary statistics values extracted from brain structure MRI data, instead of using the whole images) common in the biomedical and bio-medicine fields, deep neural networks have been less effective than their non-deep machine learning competitors, Gradient Boosting Decision Trees (GBDTs) \cite{chen2016xgboost,ke2017lightgbm,prokhorenkova2018catboost}. Nevertheless, the application of deep learning to such data is still an appealing solution \cite{arik2021tabnet,badirli2020grownet,gorishniy2021rdtl,klambauer2017snn,popov2019node} as it would enable the formation of multimodal pipelines that use additional data modalities correlated with the downstream task resulting in potentially superior performance over GBDTs.

We propose a novel incomplete multimodal learning method to predict the patients' brain disease outcomes and their cognitive status using the multimodal imaging genetics data. We train a single-modality model on MRI data and a multimodal learning model using both MRI and genetic data collaboratively. As the backbone structure of our model, we adopt an adaptation of the transformer architecture \cite{gorishniy2021rdtl}. We use a generative adversarial network (GAN) \cite{goodfellow2020gan} to guide the single-modal model to imitate the auxiliary SNP modality and enhance the main MRI modality. The empirical studies on brain imaging genetic data show our new framework improves the representational power of the model and significantly outperforms existing baselines.

\section{Method}
\subsection{Problem Formulation and Our Motivations}
In our supervised setting, we are given a dataset $D=\{x_{i}, y_{i}\}_{i=1}^{N}$, where $x_{i}=(x_{i}^{GEN}, x_{i}^{MRI}) \in X$ represents the genetic modality $x_{i}^{GEN}$ and the MRI features $x_{i}^{MRI}$ of each subject, and $y_{i}$ is the corresponding continuous subject's cognitive scores ($y_{i} \in R$) or discrete disease outcome, ($y_{i} \in \{CN,MCI,AD\}$).

\begin{figure}[!t]
    \centering
    \includegraphics[width=\textwidth]{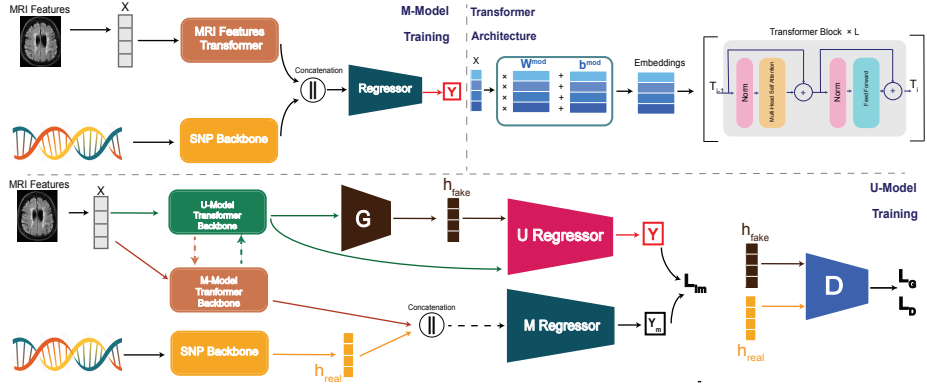}
    \caption{Overview of our incomplete multimodal framework.  \textbf{Top Left:} The $M$-Model. Neuroimaging features (measurements from 260 brain ROI volumes) and SNP data are fed into the corresponding MRI transformer component and the genetic backbone of the $M$-Model. The resulting representations are concatenated and used as the final regressor input. \textbf{Top Right:} The transformer architecture. First, 1-D features are mapped to an embedding space. Then several transformer blocks are applied. \textbf{Bottom:} Training of the $U$-Model. We guide the $U$-Model to imitate the genetic representations space of the $M$-Model using a Riemannian GAN. The MRI transformer components of the $M$-Model and the $U$-Model are trained collaboratively.}
    \label{fig:method}
\end{figure}

Fig. \ref{fig:method} illustrates our proposed incomplete multimodal learning framework. During training, we pre-train a multimodal teacher model $M$ on both MRI and genetic data, which has separate backbones for each data source. A final regressor combines the extracted features to produce the output. During testing, we only have access to MRI data. Our goal is to improve the performance of the uni-modal model $U$ that uses MRI-extracted features $\{x_{i}^{MRI}\}_{i=1}^{N}$ from brain ROI volumes to predict $y_i$. The model $U$ includes a GAN component that mimics the genetic component of the model $M$ using MRI data only. We use a transformer network as the backbone for both models $U$ and $M$. Our method substantially outperforms existing state-of-the-art GBDTs and DL networks.

\subsection{The $M$-Model}
The multimodal teacher, i.e. the $M$ Model, has two separate sub-nets designated for MRI features and SNP modalities. The choice of the subnet architecture is independent of the proposed method. Upon experiments, we show that the use of transformers outperforms other choices of MRI backbone architecture, e.g. MLP. We use a similar Transformer architecture to \cite{gorishniy2021rdtl} as the backbone. This component first takes its input $x$ to an embedding space $h$ via an affine map
\begin{equation}
    h_{j}^{mod} = x_{j}^{mod}.W_{j}^{mod} + b_{j}^{mod}
\end{equation}
with $x_{j}^{mod}$ being the $j$-th feature of a particular modality and $W_{j}^{mod}, b_{j}^{mod} \in R^{d}$ being weights and biases, respectively. The final embedding $h^{mod}$ is the simple concatenation of all embeddings of a modality:
\begin{equation}
     h^{mod} = Concat[h_{1}^{mod}, \cdots , h_{d}^{mod}]\,.
\end{equation}

Then, $L$ Transformer blocks $f_{\theta}^{k}$ are applied to the embeddings to produce the final representation of a modality $z^{mod}$:
\begin{equation}
    z^{mod} = f_{\theta}^{L} (\cdots f_{\theta}^{1}(h^{mod}))\,.
\end{equation}

Finally, a linear regression or classification head $g$ acts on the concatenation of the derived features from each modality to produce the $M$-Model output $\hat{y}$ via:
\begin{equation}
    \hat{y} = g(Concat[z^{MRI}, z^{GEN}])\,.
\end{equation}

Fig. \ref{fig:method} shows the $M$ model architecture as well as the architecture of the transformer backbone. The model is trained with the Mean Squared Error(MSE) loss.

\subsection{The $U$-Model}
The $U$-model is used for inference and shares the same MRI backbone as the $M$-model (refer to Fig. \ref{fig:method}). To enhance the $U$-model's prediction, we include a Riemannian GAN component that mimics the representation space of the genetic component of the $M$-model. By imitating the lower-dimensional representation space, our method is more efficient than attempting to estimate the entire genetic data from MRI features.

The discriminator in Riemannian GANs is different from conventional GAN models, which use Euclidean distance to distinguish between real and fake samples. Instead, the discriminator in Riemannian GANs measures the length of geodesics of a low-dimensional representation of the samples on a Riemannian manifold, such as a hypersphere. This approach provides greater stability during training because distances on a hyper-sphere are bounded, unlike Euclidean distance in the original sample space \cite{shim2020circlegan}.

To implement our Riemannian GAN, we use a trainable $d$-sphere, $S$, with a center, $c$, and a main axis, $v$, both in $R^{d}$. The discriminator takes a genetic representation, $h \in R^{d_{2}}$, whether real or fake, and maps it to an embedding, $h' \in R^{d}$, with $d < d_{2}$. It then projects $h'$ onto the $d$-sphere using $proj_{S}\mathbf{h^{i}} = \frac{h'-c}{||h'-c||_{2}}$. The discriminator assigns higher scores to samples that project onto larger radii cross-sections, with normal vector $v$. More formally, we decompose the projection of a sample $h^{i}_{proj}$ into two perpendicular components, $proj_{S}\mathbf{h^{i}} = proj_{v}\mathbf{h^{i}} + proj_{v^{\perp}}\mathbf{h^{i}}$. The discriminator then outputs the score:
\begin{equation}
D(h^{i}) = \frac{||proj_{v^{\perp}}\mathbf{h^{i}}||_{2}}{\sigma[proj_{v^{\perp}}\mathbf{h^{i}}]} -\frac{||proj_{v}\mathbf{h^{i}}||_{2}}{\sigma[proj_{v}\mathbf{h^{i}}]} \,,
\end{equation}
where $\sigma$ is the empirical standard deviation calculated over a batch of real or fake samples of size $B$, and is defined as $\sigma(h) = \sum_{i=1}^{B}\frac{||h^{i}||_{2}}{B}$.

In order to enhance stability during training and mitigate the impact of outliers and mode collapse, the GAN discriminator employs the relativistic averaged loss approach \cite{jolicoeur2019relativistic}, using the following objective function:

\begin{equation}
\begin{split}
    L_{GAN}^{D} & = - E_{h \sim p_{\text{GEN}}}[log(\text{sigmoid}(\eta(D(h) - E_{h' \sim p_{\text{U-pred}}}[D(h')])))] \\
    & -E_{h' \sim p_{\text{U-pred}}}[log(\text{sigmoid}(\eta(E_{h \sim p_{\text{GEN}}}[D(h)] - D(h'))))]
\end{split}
\end{equation}

Here, $p_{GEN}$ represents the distribution of the real samples in the genetic representation space of the M-model, $p_{U-pred}$ represents the distribution of fake samples generated by the generator module of the U-model, and $\eta$ is a hyper-parameter that adjusts the range of the score difference of the sigmoid function \cite{jolicoeur2019relativistic}.

The generator aims to generate samples that are projected near the largest hyper-section of the $d$-sphere, with an adversarial loss $L_{GAN}^{G}$ that is the inverse of the discriminator loss. The center of the $d$-sphere is identified as the point that minimizes the sum of squared distances to all samples, which is optimized using the Huber loss function \cite{huber1992robust}. This function offers a balance between mean squared error and mean absolute error, making it more resilient to outliers and suitable for a wider range of data distributions. Given a batch of samples $B=\{h^{i}\}$, we compute the following objective:
\begin{equation}
 L_{c} = \frac{1}{|B|}\sum_{i=1}^{|B|}L_{Huber}(||h_{proj}^{i} - c||_{2})
\end{equation}
Here, $L_{Huber}$ is defined as:

\begin{equation}
L_{Huber} = \begin{cases}
0.5 x^{2} & \text{if } x \leq 1 \\
x - 0.5 & \text{otherwise}
\end{cases}
\end{equation}
After fixing the center, we minimize the following objective to force the discriminator to project all samples to points with similar distances to the center and find the main axis $v$:

\begin{equation}
    L_{distance} = \frac{1}{|B|}\sum_{i=1}^{|B|}L_{Huber}(||h_{proj}^{i} - c||_{2} - \sigma[h^{i}_{proj}])
\end{equation}

We guide the U-Model to replicate the M-model representation of the genetic modality using the the Kullback-Leibler divergence between the output distributions of the U-Model and M-Model: 

\begin{equation}
    L_{im} = KL(P_{U}(y|x;T), P_{M}(y|x;T))
\end{equation}
The temperature parameter $T$ controls the distribution's sharpness. The U-Model's total loss is a weighted combination of the imitation loss $L_{im}$, the Mean Squared Error (MSE) loss $L_{MSE}$, and the U-Model generator loss $L_{GAN}^{G}$:
\begin{equation}
    L_{U} = L_{MSE} + \alpha L_{GAN} + \beta L_{im}
\end{equation}

The GAN loss is excluded from updating the backbone transformer's parameters due to its instability. We use only the MSE loss together with the imitation loss for that purpose. Also, the M-Model's MRI transformer parameters are updated using an exponential moving average scheme \cite{zhang2018deepmutual} to match the corresponding parameters in the U-Model. The update rule is expressed as:

\begin{equation}
    \theta_{M-MRI} = \gamma \theta_{M-MRI} + (1 - \gamma) \theta_{U-MRI}
\end{equation}

\section{Experiments }
The goal of our experiments is to identify how well our proposed method can predict cognitive scores and disease outcome from MRI-extracted features. We conduct experiments on our dataset and compare the proposed method with the state-of-the-art traditional machine learning and deep learning models that are suited to our data in general.
\subsection{Experimental Setup}

\subsubsection{Data Description}

In our experiments, we use the imaginge genetic data from the Alzheimer’s Disease Neuroimaging Initiative (ADNI) cohort \cite{ADNI}, which collects various datasets and biomarkers for Alzheimer's disease studies. The image features are summary statistics extracted from the structural MRIs of 844 subjects based on 260 brain ROIs \cite{christos2014Hierarchical}. 

The original ADNI cohort provides the raw cognitive scores for each patients. More recently a collaborative study by ADSP Phenotype Harmonization Consortium \cite{ADSP} conducted harmonization for the memory scores to make them comparable across different data collection institutes. Four harmonized cognitive parameters are finally provided to summarize the raw memory score sets such as MMSI, Fluency, MoCA, Trails A/B, \emph{etc}.
The harmonized cognitive parameters include four categories: the memory score (PHC\_MEM), the executive function score (PHC\_EXF), the language score (PHC\_LAN), and the visuospatial score (PHS\_VSP). However, PHS\_VSP was removed from the analysis due to a large number of missing values. 

Genetic data in our experiments are the Single Nucleotide Polymorphisms (SNP). The SNP data and all ADNI datasets had 2,001 subjects and 76K SNPs in common, with only 775 shared among all three modalities (MRI - SNP - Cognitive). These subjects comprised 175 Alzheimer's Dementia (AD), 380 Mild Cognitive Impairment (MCI), and 220 Cognitively Normal (CN) subjects. To prevent overfitting, we selected only the SNPs known to be related to AD, which were identified from a list of 287 genes with their SNPs published on the NCBI website \cite{NCBI}. After filtering out irrelevant SNPs, around 5000 SNPs were kept.

\vspace{-5pt}
\subsubsection{Genetic Data Pre-processing}

Quality Control of the SNP data consists of three main steps: removing/imputing the NA values, filtering of the SNP based on Minor Allele Frequency (MAF), and Hardy-Weinberg Equation (HWE). For this step of quality control, we removed SNPs with more than 95\% missing values and imputed the rest with the expected value as the round mean. We also filtered all the SNPs with MAF less than 0.05 and Hardy-Weinberg p-values smaller than $1\times 10^{-6}$. After this quality control, 1079 SNPs remained. 

\vspace{-5pt}
\subsubsection{Evaluation Metric and Training/Testing Settings}
We evaluate the effectiveness of our method in regressing cognitive scores using Root Mean Squared Loss (RMSE) and the coefficient of determination $r^2$. Moreover, we report macro-averaged accuracy, precision, recall and f1 score as our comparison metric for the patient disease classification task. We carry out 5-fold cross-validation (5-fold CV) to make a fair comparison of our method with the baselines. We use the Adam optimizer and L-2 regularization in training. more details about the training settings can be found on our GitHub repository when the paper is published.

\subsubsection{Baselines}

We compare with existing work in the space of structured data prediction as well as ablations that test components of our framework. Our primary points of comparison are Gradient Boosting Decision Trees (GBDT) \cite{friedman2001gbdt} variants, namely XGBoost \cite{chen2016xgboost}, LightGBM \cite{ke2017lightgbm}, and CatBoost \cite{prokhorenkova2018catboost}. However, we also evaluate other recently proposed deep learning methods Neural Oblivious Decision Ensembles (NODE) \cite{popov2019node}, GrowNet \cite{badirli2020grownet}, Self-Normalizing Networks (SNN) \cite{klambauer2017snn} as well as simple yet powerful deep architectures Transformers \cite{vaswani2017attention} and ResNet \cite{he2016resnet}.

\subsection{Experimental Results}

Table \ref{tab:regression-results} and Table \ref{tab:classification-results} show the performance of baseline models as well as our incomplete multimodal learning framework with transformer backbone on the dataset on two complex brain disorders prediction tasks: predicting cognitive scores and disease outcomes. The results demonstrate the consistent and substantial superiority of our method over the existing state of the art approaches. It must be noted that the primary point of reference of the incomplete multi-modal learning approach is the U model. This model only has access to the MRI modality. Hence the comparison with other single-modality baselines is fair.

Our method offers a twofold superiority. Firstly, the two modalities share some information that is correlated with the outcomes, and the generative adversarial network (GAN) can effectively leverage the MRI data to reconstruct the representation space of the genetic data, thereby enhancing the model's performance. Secondly, the transformer network further improves the model's representation power by identifying inter-variable dependencies that other machine learning and deep learning methods cannot detect. This results in a more accurate and robust prediction model.

\begin{table}[!t]
\centering
\caption{Comparison of performance of baselines with our proposed method. Values correspond to the mean and standard deviation of root mean squared error and $r^2$ regression coefficient across 5-fold cross-validation of regressing cognitive scores from MRI data. 
}
\label{tab:regression-results}
\resizebox{\textwidth}{!}{%
\begin{tabular}{@{}lllllll@{}}
\toprule
\multicolumn{1}{c|}{\multirow{2}{*}{Method}}                                     & \multicolumn{2}{c}{PHC\_MEM}                                                 & \multicolumn{2}{c}{PHC\_EXF}                                                 & \multicolumn{2}{c}{PHC\_LAN}                                        \\ \cmidrule(l){2-7} 
\multicolumn{1}{c|}{}                                                            & \multicolumn{1}{c}{RMSE}                & \multicolumn{1}{c|}{$r^2$}         & \multicolumn{1}{c}{RMSE}                & \multicolumn{1}{c|}{$r^2$}         & \multicolumn{1}{c}{RMSE}                & \multicolumn{1}{c}{$r^2$} \\ \midrule
\multicolumn{1}{l|}{XGBoost \cite{chen2016xgboost}}            & 0.391 ± 0.022                           & \multicolumn{1}{l|}{0.337 ± 0.071} & 0.442 ± 0.012                           & \multicolumn{1}{l|}{0.080 ± 0.073} & 0.385 ± 0.024                           & 0.192 ± 0.058             \\ \midrule
\multicolumn{1}{l|}{Light GBM \cite{ke2017lightgbm}}           & 0.377 ± 0.021                           & \multicolumn{1}{l|}{0.381 ± 0.048} & 0.419 ± 0.014                           & \multicolumn{1}{l|}{0.174 ± 0.070} & 0.385 ± 0.022                           & 0.191 ± 0.041             \\ \midrule
\multicolumn{1}{l|}{Catboost \cite{prokhorenkova2018catboost}} & 0.368 ± 0.021                           & \multicolumn{1}{l|}{0.413 ± 0.051} & 0.408 ± 0.012                           & \multicolumn{1}{l|}{0.216 ± 0.062} & 0.372 ± 0.021                           & 0.245 ± 0.045             \\ \midrule
\multicolumn{1}{l|}{MLP}                                                        & 0.367 ± 0.020                           & \multicolumn{1}{l|}{0.433 ± 0.052} & 0.376 ± 0.012                           & \multicolumn{1}{l|}{0.249 ± 0.057} & 0.327 ± 0.013                           & 0.268 ± 0.080             \\ \midrule
\multicolumn{1}{l|}{Resnet \cite{he2016resnet}}                & 0.365 ± 0.013                           & \multicolumn{1}{l|}{0.395 ± 0.056} & 0.393 ± 0.016                           & \multicolumn{1}{l|}{0.214 ± 0.082} & 0.320 ± 0.015                           & 0.271 ± 0.051             \\ \midrule
\multicolumn{1}{l|}{SNN \cite{klambauer2017snn}}               & 0.433 ± 0.018                           & \multicolumn{1}{l|}{0.358 ± 0.063} & 0.419 ± 0.029                           & \multicolumn{1}{l|}{0.349 ± 0.074} & 0.372 ± 0.036                           & 0.252 ± 0.069             \\ \midrule
\multicolumn{1}{l|}{Node \cite{popov2019node}}                 & 0.404 ± 0.033                           & \multicolumn{1}{l|}{0.388 ± 0.061} & 0.390 ± 0.027                           & \multicolumn{1}{l|}{0.237 ± 0.078} & 0.332 ± 0.022                           & 0.259 ± 0.069             \\ \midrule
\multicolumn{1}{l|}{\textbf{Our Single-Modality U-Model}}                                                  & \textbf{0.200 ± 0.090} &  \multicolumn{1}{l|}{\textbf{0.555 ± 0.051}}                                  & \textbf{0.283 ± 0.065} &      \multicolumn{1}{l|}{\textbf{0.446 ± 0.062}}                               & \textbf{0.280 ± 0.041} &           \textbf{0.301 ± 0.044}            \\ \bottomrule
\end{tabular}%
}
\end{table}

\begin{table}[!t]
\centering
\caption{Comparison of performance of baselines with our proposed method. Values correspond to the mean and standard deviation of each classification metric across 5-fold cross-validation of classifying disease progression (CN/MCI/AD) from MRI data. }
\label{tab:classification-results}
\resizebox{0.9\textwidth}{!}{%
\setlength{\tabcolsep}{20pt}
\begin{tabular}{@{}lllll@{}}
\toprule
\multicolumn{1}{c}{Method} & \multicolumn{1}{c}{Accuracy} & \multicolumn{1}{c}{Precision} & \multicolumn{1}{c}{Recall} & \multicolumn{1}{c}{F1-score} \\ \midrule
XGboost\cite{chen2016xgboost}                   & 0.790 ± 0.022                & 0.801 ± 0.023                & 0.790 ± 0.022              & 0.794 ± 0.023                \\ \midrule
Light GBM\cite{ke2017lightgbm}                  & 0.808 ± 0.026                & 0.821 ± 0.019                 & 0.808 ± 0.026              & 0.812 ± 0.024                \\ \midrule
Catboost\cite{prokhorenkova2018catboost}                   & 0.843 ± 0.031                & 0.853 ± 0.019                 & 0.843 ± 0.031              & 0.846 ± 0.033                \\ \midrule
MLP                        & 0.837 ± 0.017                & 0.835 ± 0.019                 & 0.837 ± 0.017              & 0.835 ± 0.017                \\ \midrule
Resnet\cite{he2016resnet}                     & 0.853± 0.025                 & 0.846 ± 0.023                 & 0.853 ± 0.025              & 0.848 ± 0.024                \\ \midrule
SNN\cite{klambauer2017snn}                        & 0.851 ± 0.031                & 0.860 ± 0.043                 & 0.850 ± 0.031              & 0.852 ± 0.037                \\ \midrule
Node\cite{popov2019node}                       & 0.818 ± 0.027                & 0.822 ± 0.023                 & 0.818 ± 0.027              & 0.818± 0.024                 \\ \midrule
\textbf{Ours}              & \textbf{0.927 ± 0.026}                            &   \textbf{0.925 ± 0.028}                            &     \textbf{0.927 ± 0.026}                       &     \textbf{0.925 ± 0.027}                         \\ \bottomrule
\end{tabular}%
}
\end{table}

\subsection{Ablation Studies}
We performed ablations on our proposed framework to evaluate the impact of using incomplete multimodal learning compared to a vanilla transformer. Furthermore, we examined the effectiveness of the transformer component by comparing our approach's results to an MLP backbone. Table \ref{tab:ablations} summarizes our findings, which indicate that our proposed method, in combination with the transformer backbone, yields superior results in all unimodal settings. In addition, we present the M-model's performance, which leverages both modalities for prediction and serves as an upper bound for other values.

\begin{table}[!t]
\centering
\caption{Comparison of the proposed method with ablations. Values correspond to the mean and standard deviation of root mean squared error across 5-fold cross-validation. 
}
\label{tab:ablations}
\resizebox{0.9\textwidth}{!}{%
\setlength{\tabcolsep}{22pt}
\begin{tabular}{@{}llll@{}}
\toprule
Method                                        & PHC\_MEM               & PHC\_EXF               & PHC\_LAN               \\ \midrule
Vanilla Transformer \cite{gorishniy2021rdtl}                          & 0.372 ± 0.028          & 0.361 ± 0.019          & 0.318 ± 0.016          \\ \midrule
MLP Backbone               &      0.323 ± 0.026                  &    0.359 ± 0.026                    &   0.315 ± 0.016                     \\ \midrule
M-Model (Both Modalities)                    &      0.171 ± 0.025                  &    0.247 ± 0.022                    &   0.257 ± 0.022 \\ \midrule
U-Model & 0.200 ± 0.090 & 0.283 ± 0.065 & 0.280 ± 0.041 \\ \bottomrule
\end{tabular}
}
\end{table}

\section{Conclusion}
This study introduced a novel approach to incomplete multimodal learning that leverages both neuroimaging features and genetic data to predict cognitive conditions in the brain. We validated the performance of our proposed method using ADNI imaging data and SNP genetics data and found it to outperform existing baselines. Beyond biomedical applications, our approach can be utilized in tasks where there is a primary modality and supplementary but expensive sources of data that may be available during training but not at inference.

\bibliographystyle{splncs04}
\bibliography{References/references}

\end{document}